\documentstyle[sprocl,epsfig]{article}

\def\Journal#1#2#3#4{{#1} {\bf #2}, #3 (#4)}

\def\be{\begin{equation}}
\def\ee{\end{equation}}
\def\bea{\begin{eqnarray}}
\def\eea{\end{eqnarray}}

\def\roughly#1{\mathrel{\raise.3ex\hbox{$#1$\kern-.75em%
\lower1ex\hbox{$\sim$}}}}
\def\gtrsim{\roughly>}
\def\bfp{{\bf p}}

\begin{document}

\title{
Effective Field Theory for Low-Energy $np$ Systems\footnote{´
Invited talk at the APCTP 
Workshop on ``Astro-Hadron Physics", Seoul, Korea, 
October 1997}}

\author{Tae-Sun Park\footnote{Present address: Fisica Teorica, 
Facultad de Ciencias,
Edificio de Fisicas, Universidad de Salamanca,
37008 Salamanca, Spain.}}

\address{
Department of Physics and Center for Theoretical Physics \\
 Seoul National University, Seoul 151-742, Korea\\
  E-mail: tspark@icpr.snu.ac.kr}

\maketitle\abstracts{
The properties of low-energy neutron-proton systems
are studied in an
effective field theory where only nucleons figure 
as relevant degrees of freedom.  
With a finite momentum cut-off regularization scheme,
we show that the large scattering lengths
of the $np$ systems {\em do not} spoil the convergence
of the effective field theory, 
which turns out to be
extremely successful in reproducing, with little cut-off dependence,
the deuteron properties, the $np$ $^1S_0$ scattering
amplitude and most significantly,
the $M1$ transition amplitude entering into the
radiative $np$ capture process.
}

\section{Introduction}

In this talk I would like to report on a recent work on the application
of effective field
theory to two-nucleon systems carried out 
in collaboration with Kuniharu Kubodera, Dong-Pil Min
and Mannque Rho. Effective field theories (EFTs)
have long proven to be a powerful tool 
in various areas of physics,\cite{effective,pol} 
so it is no surprise
that they are equally powerful
also in nuclear physics. 
Indeed we have recently had 
lots of successful applications of EFTs 
in low-energy nuclear dynamics.\cite{weinberg,pmr,vankolck,ksw,lm,cohen,NRZ}
One of the spectacular cases was 
the chiral perturbation theory (ChPT) calculation
of the $np\rightarrow d\gamma$ process at threshold~\cite{pmr}
with a perfect agreement with experiment.
In Ref.~\cite{pmr}, however, 
only the meson-exchange current correction
relative to the one-body $M1$ transition amplitude 
was calculated,
borrowing the latter from
the phenomenological wave function of the
accurate Argonne $v_{18}$ potential~\cite{v18}.
While there is nothing wrong there and in fact it is consistent with
the strategy of ChPT~\cite{weinberg} to calculate
only the irreducible diagrams,
there remained a ``missing link" to a complete calculation,
that is, to calculating everything within a given EFT.
The motivation of this study is to do such a
``first-principle" calculation~\cite{friar}. 
In so doing
we will compute the static properties of the bound $np$ state (the
deuteron) and the $np$ scattering amplitude in the $^1S_0$ channel.

Limiting ourselves to the processes whose typical energy-momentum 
scale is much smaller than the pion mass,
we keep only the nucleon matter field\footnote{
The anti-nucleon field is suppressed due to 
the largeness of the nucleon mass,
and is also integrated out as is in heavy-baryon formalism.}
as an explicit degree of freedom,
integrating out all massive fields as well as the pion
field.~\cite{ksw}
The resulting EFT is a non-relativistic quantum mechanics,
with all the interactions appearing as
a nucleon-nucleon potential in
Lippman-Schwinger (LS) equation.
The potential of the EFT is zero-range or contact interactions
and their derivatives, because all the meson fields which mediate
the nucleon-nucleon interactions are integrated out.
This innocent looking situation contains however
many subtleties.
First of all, even the leading order contact interactions 
are too singular to be solved by LS equation
in three-dimensional space, 
thus we need to introduce
a regulator and a renormalization scheme to handle the singularity.
While the appearance of such a singularity is quite common
in quantum field theories,
the real subtlety comes from the fact that
the $np$ states in nature are all very close
to the threshold: they are either weakly bound ($^3S_1$) or almost
bound ($^1S_0$). Those states near threshold 
cannot be treated by perturbation expansion at all,
which means that all the
``reducible diagrams" up to infinite order should be summed by
solving LS (or Schr\"odinger) equation.
Furthermore, those states have huge scattering
lengths $a$, and the appearance of extremely small mass scales
$a^{-1}$ makes the convergence of EFTs by no means trivial.
Indeed, using the dimensional regularization
which has  proven to be a very convenient and successful tool
in handling singularities in most cases,
Kaplan, Savage and Wise
\cite{ksw} and Luke and Manohar \cite{lm} have shown that
the EFT breaks down at a very small scale, $p_{crit} = \sqrt{\frac{2}{a
r_e}}$ for large scattering length $a$, where $r_e$ is the
effective range and that this problem cannot be ameliorated by
introducing the pionic degree of freedom. 
It was followed by the observation 
by Beane et al \cite{cohen} that,
for nonperturbative cases such as ours,
the physical results after the renormalization procedure
may still depend on the regularization scheme.
As pointed out by 
Beane et al \cite{cohen} 
and Lepage,\cite{lepage} the problem can
however be resolved if one uses a cut-off regularization. 
In EFTs, the cut-off has a physical meaning and hence 
should be set by the mass of the lightest degree of freedom
which is integrated out, namely the pion in our case.
If one chooses too low a cut-off, 
the valid region of EFTs unnecessarily shrinks down,
while if one chooses too high a cut-off, one
introduces irrelevant degrees of freedom and hence makes the theory
unnecessarily complicated.
We find that the optimal cut-off in our case is 
$\Lambda \sim 200$ MeV as one can see from the results in 
Table~\ref{table:result} and Figures 1 and 2.

\section{Renormalization and Phase shift}

To do the calculation algebraically, 
we choose the following form of
regularization appropriate to a separable potential given by the
local Lagrangian:
\begin{equation}
\langle {\bf p'} | {\hat V} | {\bf p} \rangle
= S_\Lambda({{\bf p}'}^2) V({\bf p}'-{\bf p}) S_\Lambda({\bf p}^2)
\label{V}\end{equation}
where $S_\Lambda({\bf p}^2)$ 
is a regulator which suppresses the contributions from $|{\bf p}|
\gtrsim
\Lambda$, $\lim_{|\bfp| \ll \Lambda} S_\Lambda(\bfp^2) =1$ 
and $\lim_{|\bfp|\gg \Lambda} S_\Lambda(\bfp^2)=0$, 
and $V({\bf q})$ is a finite-order polynomial in ${\bf q}$. 
We shall do the calculation to the next-to-leading order (NLO),
and so the most general form of $V({\bf q})$ is
\begin{equation}
V({\bf q}) = \frac{4\pi}{M} \left(C_0 +
(C_2 \delta^{ij} + D_2 \sigma^{ij}) q^i q^j \right),
\label{Vq}\end{equation}
where $M$ is the nucleon mass and
$\sigma^{ij}$ is the rank-two tensor that is effective only in
the spin-triplet channel,
\begin{equation}
\sigma^{ij} = \frac{3}{\sqrt{8}} \left(
\frac{\sigma_1^i \sigma_2^j + \sigma_1^j \sigma_2^i}{2}
- \frac{\delta^{ij}}{3} \sigma_1 \cdot \sigma_2 \right).
\end{equation}
Note that the coefficients $C_{0,2}$ are (spin) channel-dependent,
and that $D_2$ is effective only in spin-triplet channel.
Thus we have five parameters; two in $^1S_0$ and three in $^3S_1$ channel,
which will be fixed from experiments. 
Since the explicit form of the regulator should not
matter,\cite{lepage} we shall choose the Gaussian form,
\begin{equation}
S_\Lambda({\bf p}^2) = \exp\left(- \frac{{\bf
p}^2}{2\Lambda^2}\right)
\label{S}\end{equation}
where $\Lambda$ is the cut-off. 
The LS equation for the wavefunction
$|\psi\rangle$,
$
| \psi\rangle = | \varphi\rangle
+ {\hat G}^0 \,{\hat V} | \psi\rangle
$
where $|\varphi\rangle$ is the free wavefunction and ${\hat G}^0$
is the free two-nucleon propagator depending on the total energy
$E$, $
\langle {\bf p}' | {\hat G}^0 | {\bf p} \rangle
 = \frac{\langle {\bf p}' | {\bf p}\rangle}{
      E - \frac{{\bf p}^2}{M} + i 0^+}$
leads to
the $S$-wave function (for the potential (\ref{Vq})) of the form
\begin{eqnarray}
\psi({\bf r}) &=& \varphi({\bf r})
 + \frac{S_\Lambda(M E)\,C_E}{1- \Gamma_E C_E} \,
  \left[
1 - \frac{\sqrt{Z} C_2}{C_E} (\nabla^2 + ME)
\right.
\nonumber \\
&&\left.
- \frac{\sqrt{Z} D_2}{C_E}
 \frac{S_{12}({\hat r})}{\sqrt{8}}
 r \frac{\partial}{\partial r} \frac{1}{r} \frac{\partial}{\partial r}\right]
 {\tilde \Gamma}_\Lambda({\bf r})
\label{fullwave}
\end{eqnarray}
where $S_{12}(\hat r)= 3 \sigma_1\cdot \hat r \,\sigma_2\cdot \hat r
 - \sigma_1\cdot \sigma_2$,
\begin{eqnarray}
\Gamma_E &=&
4\pi \int \frac{d^3{\bf p}}{(2\pi)^3}
\frac{S_\Lambda^2({\bf p}^2)}{ME - {\bf p}^2 + i0^+},
\label{GE}
\\
{\tilde \Gamma}_\Lambda({\bf r}) &=&
4\pi \int \frac{d^3{\bf p}}{(2\pi)^3} \frac{S_\Lambda({\bf p}^2)}{ME - {\bf p}^2
 + i0^+} \mbox{e}^{i {\bf p}\cdot {\bf r}}\label{GLambda} ,
\\
\frac{1}{\sqrt{Z}} &=& 1 - C_2 I_2,
\label{Z}\\
C_E &=& a_\Lambda \left(1 + \frac12 a_\Lambda r_\Lambda ME\right)
 + \left(\sqrt{Z} D_2 ME\right)^2 \Gamma_E ,
\end{eqnarray}
with
\begin{eqnarray}
a_\Lambda &\equiv& Z\left[C_0 + (C_2^2 + \delta_{S,1} D_2^2) I_4)\right],
\\
r_\Lambda  &\equiv& \frac{2 Z}{a_\Lambda^2} \left[
   2 C_2 - (C_2^2 - \delta_{S,1} D_2^2) I_2 \right]
\label{arC}\end{eqnarray}
where $I_{n}$ ($n=2,\,4$) are defined by
\begin{equation}
I_{n} \equiv
 - \frac{\Lambda^{n+1}}{\pi} \int_{-\infty}^\infty
 dx\, x^{n} S_\Lambda^2(x^2 \Lambda^2).
\end{equation}
With the regulator (\ref{S}), the integrals come out to be $I_2=
-\frac{1}{2\sqrt{\pi}}
\Lambda^3$ and $I_4= -\frac{3}{4\sqrt{\pi}} \Lambda^5$.

The phase shifts can be calculated by looking at the large-$r$
behavior of the wavefunction. To do this, it is convenient to
separate the pole contributions of the integrals Eqs.(\ref{GE},
\ref{GLambda}) as
\begin{eqnarray}
\Gamma_E &=&  - i \sqrt{ME}\, S_\Lambda^2(ME)
 + I_\Lambda(E),
\\
{\tilde \Gamma}_\Lambda({\bf r}) &=&
-\frac{S_\Lambda(ME)}{r} \left[
 \mbox{e}^{i \sqrt{ME} r } - H(\Lambda r, \frac{ME}{\Lambda^2})
\right],
\end{eqnarray}
which define the functions $I_\Lambda(E)= \Lambda
I(\frac{ME}{\Lambda^2})$ and $H(\Lambda r, \frac{ME}{\Lambda^2})$,
both of which are real.  Note that $H(0,\,\varepsilon)=1$ which
makes ${\tilde \Gamma}_\Lambda({\bf 0})$ finite, and that
$\lim_{x\gg 1} H(x,\,\varepsilon)=0$. 
The $^1S_0$ phase shift $\delta(^1S_0)$
takes the form
\begin{equation}
p\cot \delta(^1S_0) = \frac{1}{S_\Lambda^2(ME)} \left[
 I_\Lambda(E)
 - \frac{1}{a_\Lambda (1 + \frac12 a_\Lambda r_\Lambda ME)}
\right].
\label{pCot1S0}\end{equation}
The ``effective" low-energy constants, $a_\Lambda$ and $r_\Lambda$,
are fixed by comparing (\ref{pCot1S0}) to the 
{\it effective-range expansion}
\begin{equation}
p\cot\delta = -\frac{1}{a} + \frac12 r_e p^2 + \cdots ,
\label{ERT}
\end{equation}
\begin{eqnarray}
\frac{1}{a_\Lambda} &=& \frac{1}{a} + \Lambda I(0)
 = \frac{1}{a} - \frac{\Lambda}{\sqrt{\pi}},
\\
r_\Lambda &=& r_e - \frac{2 I'(0)}{\Lambda} - 
 \frac{4}{a} \left[\frac{\partial}{\partial \bfp^2} S_\Lambda(\bfp^2)
 \right]_{\bfp^2=0}
= r_e - \frac{4}{\sqrt{\pi}\Lambda} + \frac{2}{a \Lambda^2}.
\end{eqnarray}
They then give us 
the ``renormalization conditions" 
of the $C_0$ and $C_2$, 
with a given value of $\Lambda$.
Two important observations to
make here: (a)  We note that there is an upper bound of $\Lambda$,
$\Lambda_{\rm Max}$, if one requires that $Z$ be positive and that
$C_2$ be real. That is, 
for $\Lambda >\Lambda_{\rm Max}$, the potential of the EFT becomes
non-Hermitian. With $a= -23.732$ fm and $r_e= 2.697$ fm for the $^1S_0$
channel taken from the Argonne $v_{18}$ potential \cite{v18} (which
we take to be ``experimental"), we find that $\Lambda_{\rm Max}
\simeq 348.0$ MeV; (b)
the value $\Lambda_{Z=1}$ defined such that $Z=1$ when
$\Lambda=\Lambda_{Z=1} \simeq 172.2$ MeV is  quite special. At this
point, we have $r_\Lambda=0$ and $C_2=0$, that is, the NLO
contribution is identically zero. This corresponds to the
leading-order calculation with the $\Lambda$ chosen to fit the
experimental value of the effective range $r_e$.
A similar observation was made by Beane et al \cite{cohen} using
a square-well potential in coordinate space with a radius $R$, 
with $R^{-1}$ playing the role of $\Lambda$.

The resulting phase shift with $\Lambda=\Lambda_{Z=1}$ is plotted
in Fig.\ 1. We see that the agreement with the result taken from
the  Argonne $v_{18}$ potential \cite{v18} is perfect up to 
$p \sim m_\pi/2$. Beyond that, we should expect corrections from
the next-to-next-order and higher-order terms. 
In Fig.\ 2, we show
how the phase-shift for a fixed center-of-mass momentum, 
$p= 68.5$ MeV varies as the cut-off is changed. 
The solid curve is our NLO result, 
the dotted one the LO result (with $C_2=0$), 
and the horizontal dashed line the result taken 
from the $v_{18}$ potential (``experimental"). 
We find that our NLO result is
remarkably insensitive to the value of $\Lambda$ for 
$\Lambda\gtrsim m_\pi$.
It demonstrates that,
going to the higher-order calculations in EFTs,
we have not only more accurate agreement with the data
but also less dependence on the cut-off.

\begin{figure}[htbp]
\centerline{\epsfig{file=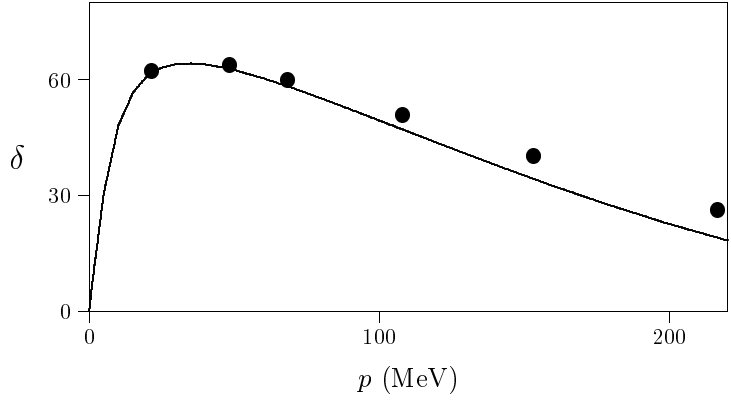,width=3.75in}}
\caption[phase]{\protect \small
$np$ $^1S_0$ phase shift (degrees) vs. 
the center-of-mass (CM) momentum $p$. 
Our theory with $\Lambda=\Lambda_{Z=1}\simeq
172$ MeV is given by the solid line, and the results from the
Argonne $v_{18}$ potential \cite{v18} (``experiments") by the solid
dots.}
\end{figure}

\begin{figure}[htbp]
\centerline{\epsfig{file=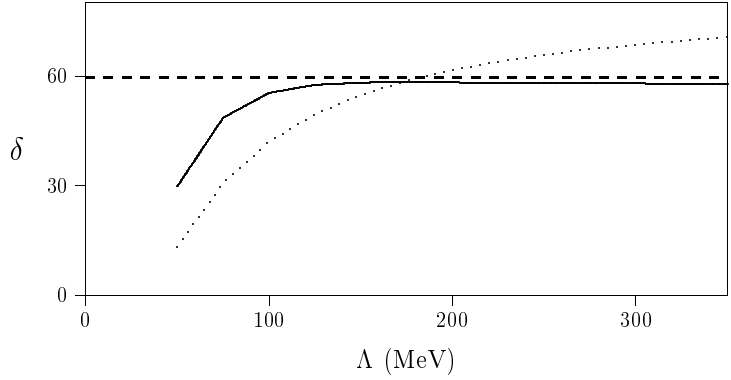,width=3.75in}}
\caption[deviate]{\protect \small
$np$ $^1 S_0$ phase shift (degrees) vs. the cut-off $\Lambda$ for a
fixed CM momentum $p= 68.5$ MeV. 
The solid curve represents the
NLO result, the dotted curve
 the LO result and the horizontal dashed line the
result from the $v_{18}$ potential \cite{v18}.}
\end{figure}
\begin{table}
\caption{$C_0$, $C_2$ and $D_2$
for various values of $\Lambda$.
The unit of $C_0$ is $\mbox{GeV}^{-1}$
while that of $C_2$ and $D_2$ is $\mbox{GeV}^{-3}$.
$D_2$ in $^1S_0$ channel are identically zero.}\label{coeff}
\vspace{0.1cm}
\begin{center}
\begin{tabular}{|c||rr|rrr|} \hline
{} & \multicolumn{2}{|c|}{$^1S_0$} &\multicolumn{3}{|c|}{$^3S_1$} 
\\ \cline{2-6}
$\Lambda$ (MeV) & $C_0$ & $C_2$ & 
  $C_0$ & $C_2$ & $D_2$ \\
\hline
$150.0$ & $-9.877$ & $-56.4 $ & $-15.503$ & $-206.6$ & $186.9$
\\ 
$172.2$ & $-9.482$ & $0$ & $-12.051$ & $-78.0$ & $169.7$
\\
$200.0$ & $-10.185$ & $37.9 $ & $-9.925$ & $2.5$ & $156.3$
\\
$216.1$ & $-11.377$ & $51.8$ & $-9.730$ & $31.7$ & $152.7$
\\ 
$250.0$ & $-17.167$ &  $74.2$ & $-14.342$ &$83.4$ & $158.1$
\\ \hline 
\end{tabular}
\end{center}
\end{table}

As for the $^3S_1$ coupled channel, the
phase shift $^3S_1$ and the mixing angle $\epsilon_1$
are given as
\bea
p\cot \delta(^3S_1) &=&
\frac{1}{S_\Lambda^2(ME)} \left[
 I_\Lambda(E)
 - \frac{1-\eta^2(E)}{a_\Lambda (1 + \frac12 a_\Lambda r_\Lambda ME)}
\right],
\\
\frac{\eta(E)}{1- \eta^2(E)} &=&
\frac{\sqrt{Z} D_2  ME}{a_\Lambda\left(1 + \frac12 a_\Lambda r_\Lambda ME
\right)},
\label{etaE}
\eea
where $\eta(E) \equiv -\tan \epsilon_1$ 
and we have used the eigenphase parametrization.\cite{eigenphase}
The $D_2$ can be fixed by the deuteron $D/S$
ratio $\eta_d\simeq 0.025$ \cite{v18} at 
$E= -B_d$ with $B_d$ the binding energy of the deuteron,
\begin{equation}
\sqrt{Z} D_2 = \frac{\eta_d}{1-\eta_d^2}
 \frac{a_\Lambda}{-M B_d}
 \left[ 1 - \frac{1}{2} a_\Lambda r_\Lambda M B_d
 \right].
\end{equation}
The renormalization procedure is
the same as for the $^1S_0$ channel. 
The only difference is that
the value of $\Lambda_{Z=1}$ that makes $Z=1$ does not coincide
with $\Lambda_{r_\Lambda=0}$ that makes $r_\Lambda=0$. 
Using $a= 5.419$ fm and $r_e= 1.753$ fm \cite{v18} 
for the $^3S_1$ channel, we find
that $\Lambda_{\rm Max} = 304.0$ MeV, $\Lambda_{Z=1} = 198.8$ MeV and
$\Lambda_{r_\Lambda=0}= 216.1$ MeV. 
The values of $C_0$, $C_2$ and $D_2$ 
with respect to various values of $\Lambda$
are listed in Table~\ref{coeff}.
One can see the ``naturalness" of the coefficient
by making them dimensionless, for example,
$m_\pi C_0$, $m_\pi^3 C_2$ and $m_\pi^3 D_2$
in spin-triplet channel with $\Lambda=216.1$ MeV are
$-1.4$, $0.1$ and $0.4$, respectively.

\section{Results and Discussion}

Given $C_0$, $C_2$ and $D_2$ for a given $\Lambda$, all other
quantities are predictions. The binding energy of the deuteron is
determined by the pole position,
\begin{equation}
\gamma S_\Lambda^2(-\gamma^2) + I_\Lambda(-\gamma^2)
= \frac{1-\eta^2_d}{
  a_\Lambda \left(1 - \frac12 a_\Lambda r_\Lambda \gamma^2
 \right) }
\end{equation}
with $\gamma\equiv \sqrt{M B_d}$. 
The $S$- and $D$-wave radial
wavefunctions of the deuteron are
\begin{eqnarray}
u(r) &=& \mbox{e}^{-\gamma r} - H(\Lambda r, \frac{-\gamma^2}{\Lambda^2})
 + \beta_\Lambda \frac{4\pi r}{\Lambda^2} \delta_\Lambda^{(3)}({\bf r}),
\\
\omega(r) &=& \eta_d \frac{r^2}{\gamma^2}
 \frac{\partial}{\partial r} \frac{1}{r} \frac{\partial}{\partial r}
 \frac{1}{r} \left[
 \mbox{e}^{-\gamma r} - H(\Lambda r, \frac{-\gamma^2}{\Lambda^2})
 \right],
\end{eqnarray}
where 
\begin{eqnarray}
\delta^{(3)}_\Lambda({\bf r}) &=&
 \int \frac{d^3{\bf p}}{(2\pi)^3} S_\Lambda({\bf p}^2)
 \mbox{e}^{i {\bf p}\cdot {\bf r}},
\\
\beta_\Lambda &=& \frac{(\sqrt{Z}-1) \Lambda^2}{
 a_\Lambda(1 + \frac12 a_\Lambda r_\Lambda ME) I_2
 S_\Lambda(ME)} .
\end{eqnarray}

We now have all the machinery to calculate the deuteron properties:
the wavefunction normalization factor $A_s$, the radius $r_d$, the
quadrupole moment $Q_d$ and the $D$-state probability $P_D$,
which are defined as
\bea
A_s^{-2} &=& \int_0^\infty\! dr\, \left[
 u^2(r) + \omega^2(r) \right],
\nonumber \\
r_d^2 &=& \frac{A_s^2}{4}\,\int_0^\infty\! dr\, r^2\,\left[
 u^2(r) + \omega^2(r) \right],
\nonumber \\
Q_d &=& \frac{A_s^2}{\sqrt{50}}\,\int_0^\infty\! dr\, r^2\,\left[
 u(r)\omega(r)  - \frac{\omega^2(r)}{\sqrt{8}}\right],
\nonumber \\
P_D &=& A_s^2\,\int_0^\infty\! dr\, \omega^2(r) .
\eea
The magnetic moment of the deuteron $\mu_d$ is related 
to the $P_D$ through
\begin{equation}
\mu_d= \mu_S - \frac32 \left(\mu_S -\frac12\right) P_D
\label{mud}
\end{equation}
where $\mu_S \simeq 0.8798$ is the isoscalar
nucleon magnetic moment. Finally the one-body isovector $M1$
transition amplitude  relevant for $n+p \rightarrow d +
\gamma$ at threshold~\cite{pmr} is
\begin{equation}
M_{\rm 1B}\equiv \int_0^\infty\! dr\, u(r) u_0(r)
\end{equation}
where $u_0(r)$ is the $np$ $^1S_0$
radial function,
\begin{eqnarray}
u_0(r) &=& \frac{\sin(\sqrt{ME}\, r + \delta({}^1S_0))}{\sin\delta({}^1S_0)}
 - H(\Lambda r, \frac{ME}{\Lambda^2})
+ \beta_\Lambda \frac{4 \pi r}{\Lambda^2} \delta^{(3)}_\Lambda({\bf r})
\end{eqnarray}

\begin{table}
\caption[deuteronTable]{Deuteron properties and
the $M1$ transition amplitude entering into the $np$ capture for
various values of $\Lambda$.}\label{table:result}
\begin{center}
\begin{tabular}{|c|cccc|cc|} \hline
$\Lambda$ (MeV) & $150$ & $198.8$ & $216.1$ & $250$ & Exp. & $v_{18}$\cite{v18} \\
\hline
$B_d$ (MeV) & $1.799$ & $2.114$ & $2.211$ & $2.389$ & $2.225$ & 2.225\\
$A_s$ ($\mbox{fm}^{-\frac12}$)
   & 0.869 & 0.877 & 0.878 & 0.878 & 0.8846(8) & 0.885 \\
$r_d$ (fm)
   & 1.951 & 1.960 & 1.963 & 1.969 & 1.966(7) & 1.967 \\
$Q_d$ ($\mbox{fm}^2$)
   & 0.231 & 0.277 & 0.288 & 0.305 & 0.286 & 0.270 \\
$P_D$ (\%)
   & 2.11 & 4.61 & 5.89 & 9.09 & $-$ & 5.76 \\
$\mu_d$
   & 0.868 & 0.854 & 0.846 & 0.828 & 0.8574 & 0.847 \\
$M_{\rm 1B}$ (fm)
   & 4.06 & 4.01 & 3.99 & 3.96 & $-$ & 3.98 
\\ \hline
\end{tabular}
\end{center}
\end{table}

The (parameter-free) numerical results  are listed 
in Table~\ref{table:result} for
various values of the cut-off $\Lambda$. We see that the agreement
with the experiments (particularly for $\Lambda=216.1$ MeV)
is excellent with very little dependence on
the precise value of $\Lambda$. 
It may be coincidental but highly 
remarkable that even the quadrupole moment, 
which (as the authors of Ref.~\cite{v18} stressed) 
the $v_{18}$ potential fails to reproduce, comes out correctly.

Let us compare our result (\ref{pCot1S0}) with that
obtained with the dimensional regularization~\cite{ksw},
\begin{equation}
\left. p\cot\delta\right|_{Dim.} 
= - \frac{1}{a (1 + \frac12 a r_e ME)}.
\label{dimension}
\end{equation}
Expanding $p\cot\delta$ of (\ref{dimension}) in $ME$, we find that
the coefficient of the $n$-th order term is order of $a^{n-1}
r_e^n$. This increases rapidly with $n$ when $a$ is large, disagreeing
strongly with the fact that the low-energy
scattering is well described by just two terms of the {\it
effective range expansion} in (\ref{ERT}). This observation led the
authors of Ref.~\cite{ksw} to conclude that the critical momentum scale
at which the EFT expansion breaks down is very small for a very large
$a$:
\begin{equation}
\left. p_{crit}\right|_{Dim} \sim \sqrt{\frac{2}{a r_e}}.
\end{equation}
We arrive at a different conclusion. With the cut-off
regularization, the scattering length $a$ is replaced by 
an {\em effective one}, $a_\Lambda$, that is
order of $\Lambda^{-1}$ for large $|a|$. 
This agrees with the findings of
Beane et al \cite{cohen} and Lepage.\cite{lepage}
Counting $r_e$ to be order of $\Lambda^{-1}$, the $n$-th order
coefficient now is $\Lambda^{1-2n}$, as one would expect on a
general ground.
Recalling that a small (and negative) scattering length
corresponds to weak (and attractive) interactions,
it is also remarkable that
the mechanism of the finite cut-off EFTs is quite similar 
to the ``quasi-particle" phenomena
since both convert highly non-linear systems
into weakly interacting systems.

Using the finite cut-off regularization scheme
which is the most faithful way to realize the 
principles of EFTs,
we have demonstrated that
the low-energy nuclear physics can be well-described
by EFTs.
In particular, it is satisfying that 
the classic $np$ capture process can be completely
understood from a ``first-principle" approach. 
Here the cut-off regularization was found to be highly efficient.
With the dimensional regularization 
the $M1$ matrix element was found to be in total disagreement with
the result of the Argonne $v_{18}$ potential. 

There are several important and urgent
extensions and applications of the finite EFTs in nuclear physics.
One of the most important applications is the proton fusion process,
$p+p\rightarrow d+e^+ + \nu$, at threshold
which plays a crucial role in the stellar evolution.\cite{bahcall}
This process has been recently worked out and will 
appear soon.\cite{PKMR}
{}From the theoretical side, the most urgent task is
to take into account the pion degree of freedom so as to extend the calculation
to higher chiral order. 
This would enable us to study the interplay
between the breakdown of an EFT and the emergence of a ``new physics."

\section*{Acknowledgments}
It is an honor and pleasure for me  to attend
this Workshop organized to celebrate 
Prof. Rho's 60th birthday and
to present the work in which he actively participated.
I would like to express my sincere gratitude to him
for his advice, encouragements and efforts extended to me
since my graduate school.
This work was supported in part
by the Korea Science and Engineering Foundation
through CTP of SNU
and in part by the Korea Ministry of Education under 
the grant BSRI-97-2441.

\end{document}